\documentclass[aps,prl,twocolumn,superscriptaddress]{revtex4-2}

\usepackage{gensymb}
\usepackage{graphicx}
\usepackage{dcolumn}
\usepackage{bm}
\usepackage[colorlinks,linkcolor=blue,anchorcolor=blue,citecolor=blue,urlcolor=blue,filecolor=blue,menucolor=blue,runcolor=blue]{hyperref}
\usepackage{ulem}

\begin{document}

\title{Tuning a magnetic energy scale with pressure in UTe$_2$}

\author{Hyunsoo Kim}
\affiliation{Department of Physics, Missouri University of Science and Technology, Rolla, MO 65409, USA}
\affiliation{Maryland Quantum Materials Center and Department of Physics, University of Maryland, College Park, Maryland, USA}

\author{I-Lin Liu}
\affiliation{Maryland Quantum Materials Center and Department of Physics, University of Maryland, College Park, Maryland, USA}

\author{Wen-Chen Lin}
\affiliation{Maryland Quantum Materials Center and Department of Physics, University of Maryland, College Park, Maryland, USA}

\author{Yun Suk Eo}
\affiliation{Maryland Quantum Materials Center and Department of Physics, University of Maryland, College Park, Maryland, USA}

\author{Sheng Ran}
\affiliation{Maryland Quantum Materials Center and Department of Physics, University of Maryland, College Park, Maryland, USA}

\author{Nicholas P. Butch}
\affiliation{Maryland Quantum Materials Center and Department of Physics, University of Maryland, College Park, Maryland, USA}
\affiliation{NIST Center for Neutron Research, National Institute of Standards and Technology, Gaithersburg, MD 20899, USA}

\author{Johnpierre Paglione}
\affiliation{Maryland Quantum Materials Center and Department of Physics, University of Maryland, College Park, Maryland, USA}
\affiliation{Canadian Institute for Advanced Research, Toronto, Ontario M5G 1Z8, Canada}
\email{paglione@umd.edu}

\date{\today}

\begin{abstract}
A fragile ordered state can be easily tuned by various external parameters. When the ordered state is suppressed to zero temperature, a quantum phase transition occurs, which is often marked by the appearance of unconventional superconductivity. While the quantum critical point can be hidden, the influence of the quantum criticality extends to fairly high temperatures, manifesting the non-Fermi liquid behavior in the wide range of the $p$-$H$-$T$ phase space. Here, we report the tuning of a magnetic energy scale in the heavy-fermion superconductor UTe$_2$, previously identified as a peak in the $c$-axis electrical transport, with applied hydrostatic pressure and magnetic field along the $a$-axis as complementary (and opposing) tuning parameters. Upon increasing pressure, the characteristic $c$-axis peak moves to a lower temperature before vanishing near the critical pressure of about 15 kbar. The application of a magnetic field broadens the peak under all studied pressure values. The observed Fermi-liquid behavior at ambient pressure is violated near the critical pressure, exhibiting nearly linear resistivity in temperature and an enhanced pre-factor. Our results provide a clear picture of energy scale evolution relevant to magnetic quantum criticality in UTe$_2$.
\end{abstract}



\maketitle

Few systems in nature exhibit a fragile long-range magnetic order, where the thermal phase transition into its ordered state can be readily suppressed by either chemical substitution, magnetic field, or physical pressure. However, systems have been found that undergo a quantum phase transition at a critical value of the tuning parameter \cite{Mathur1998, Paglione2003, Stockert2011, Shibauchi2014}, deemed a quantum critical point (QCP). However, the QCP is often putative, being hidden within a surrounding superconducting phase which is thought to be mediated by fluctuations affiliated with the magnetic order \cite{Mathur1998}.
While the majority of magnetic unconventional superconductors are found near an antiferromagnetic instability, several uranium-based superconductors including URhGe and UCoGe coexist with ferromagnetism \cite{Aoki2019review}, making them promising candidates for a topological spin-triplet superconductivity \cite{Ran2019science}.

Recently, UTe$_2$ was identified as a new member of the U-based superconductor family \cite{Ran2019science}, with a transition temperature $T_c$ reaching up to 2~K \cite{Rosa2021}.
The normal state of UTe$_2$ can be described by the Kondo lattice model where the localized magnetic moment of uranium is hybridized with the conduction electrons at low temperatures \cite{Kang2022}.
UTe$_2$ does not magnetically order, but the superconductivity in this paramagnetic heavy fermion is believed to be in the vicinity of the magnetic instability \cite{Ran2019science}.
The application of pressure as low as 15 kbar induces a long-range magnetic order \cite{Ran2020}.
Because of the relatively small energy scales of the superconductivity and magnetic order in UTe$_2$, a rich phase diagram emerges when the system is subjected to external parameters. 
However, understanding of competition and interplay between magnetism and superconductivity in UTe$_2$ remains elusive, and the associated quantum criticality in the $p$-$H$-$T$ phase space has not been fully explored.

In UTe$_2$, electrical resistivity exhibits the behavior of a Fermi liquid in its temperature dependence above $T_c$ for currents applied along all three crystallographic axes \cite{Eo2022}.
Whereas the resistivity along the $a$ and $b$ directions is consistent with typical incoherent-to-coherent crossover upon cooling as expected for a Kondo lattice at low temperatures, Eo {\it et al.} \cite{Eo2022} found a qualitatively different behavior in the $c$-axis transport which exhibits a pronounced local maximum near 12 K.
An anomaly in $d\rho_a/dT$ \cite{Willa2021, Eo2022} and $\chi_a$ (magnetic susceptibility with a field along the $a$-axis) \cite{Li2021} was reported at the same temperature.
The pressure evolution of $\chi_a$ was studied by Li {\it et al.} \cite{Li2021} where the feature moves to lower temperatures with pressure.
In contrast, $\chi_b$ exhibit a broad local maximum around 35 K, and its pressure evolution is scaled with that of the metamagnetic transition field \cite{Knebel2020}.
A similar peak in the electrical transport measurement was identified at 16 K in an unoriented sample \cite{Cairns2020} and in a sample under applied pressure \cite{Ran2020}. 
Other measurements including heat capacity \cite{Willa2021}, linear thermal expansion coefficient \cite{Thomas2021} and thermoelectric power \cite{Niu2020prl} exhibit a prominent feature around 12 K.
The $c$-axis peak has been associated with a spin fluctuation energy scale based on thermodynamic measurements \cite{Willa2021}, and therefore a measurement of $c$-axis transport as a function of tunable parameters allows for direct tracking of the evolution of this energy scale and any resultant change in physical properties, providing a straightforward but crucial window into the magnetic fluctuation spectrum likely responsible for superconductivity in UTe$_2$. 

\begin{figure*}
\includegraphics[width=1\linewidth]{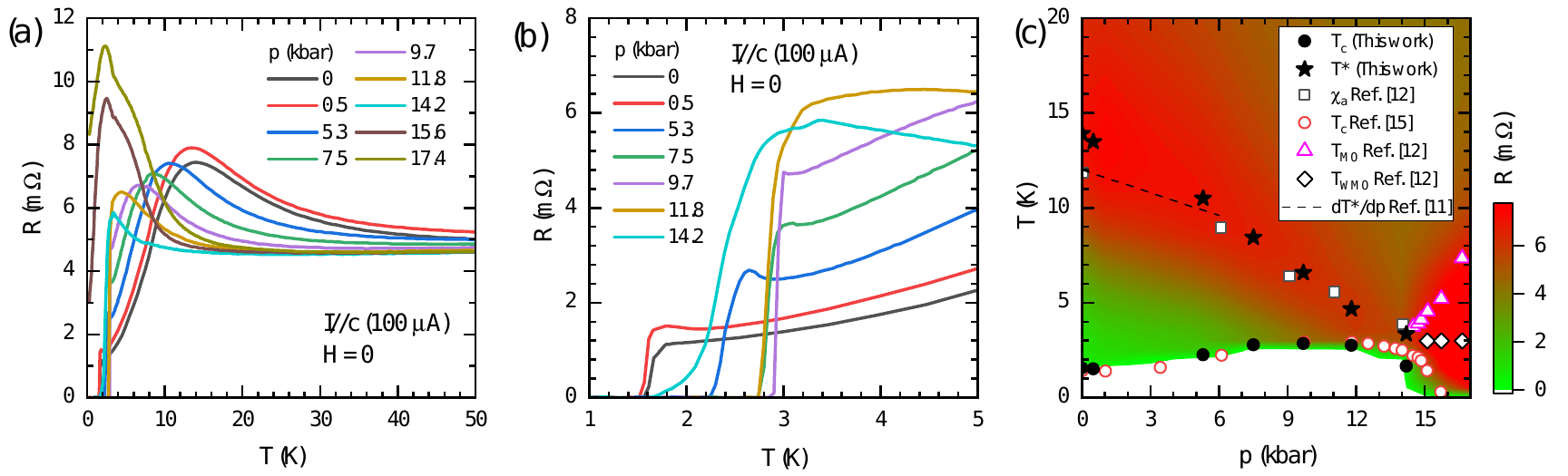}%
\caption{\label{fig1} 
{\bf Pressure evolution of the $c$-axis resistivity of UTe$_2$ in the absence of a magnetic field.} 
Panel (a) shows resistance $R$ of UTe$_2$ measured with the electrical current applied along the crystallographic $c$-axis under various applied pressures up to $p=17.4$ kbar.
The peak in $R(T)$ monotonically moves towards the lower temperature with increasing pressure.
The pressure evolution of the resistive superconducting transition is shown in panel (b) for pressures up to 14.2 kbar, above which zero resistance is not observed. 
Panel (c) exhibits a phase diagram of the characteristic temperature scales (various symbols) of the system overlaid on a color contour presentation of the resistance $R$ variation with pressure and temperature. 
The black (this work) and red \cite{Thomas2021} circles represent the superconducting transition and the black squares indicate a shoulder-like feature observed in magnetic susceptibility $\chi_a$ \cite{Li2021}, which closely tracks the position of the maximum in $c$-axis resistivity ($T^*$) plotted as black stars.
The dashed line represent the suppression of the observed minimum of the thermal expansion coefficient, estimated by using a thermodynamic relationship of electronic Gr\"uneisen parameter \cite{Willa2021}, and the triangle and diamond symbols observed above 14.2 kbar are features attributed to magnetic ordering \cite{Li2021,Thomas2021}.} 
\end{figure*}

In this work, we investigate the $c$-axis electrical transport in UTe$_2$ while tuning the applied magnetic field and pressure in order to elucidate the presence of quantum criticality in its rich phase diagram.
By performing precision measurements of the electrical resistance $R$ under applied pressures up to 17.4 kbar and in magnetic fields up to 18 T applied along the $a$-axis, we determine the pressure and field evolution of the characteristic fluctuation energy scale, upper critical field, and the power-law behavior of the $c$-axis electrical resistance. Our results clearly indicate an energy scale evolution relevant to magnetic quantum criticality in UTe$_2$.

Figure \ref{fig1} presents the applied pressure dependence of $R(T)$ in UTe$_2$ with electrical currents applied along the crystallographic $c$-axis.
The measured single-crystal sample was grown by the chemical vapor transport method, and achieves zero resistance at $T_c$=1.6 K in the absence of pressure (see Methods for detail). 
The ambient pressure ($0$ kbar) $R(T)$ curve exhibits the characteristic $c$-axis peak near 13 K as shown previously \cite{Eo2022}, which monotonically moves towards lower temperatures with increasing applied pressures while $T_c$ steadily increases as reported previously \cite{Thomas2020, Aoki2020}, reaching a maximum at $p=9.7$ kbar before decreasing rapidly.
The resistive superconducting transition itself exhibits distinct features that evolve with pressure as shown in Fig. \ref{fig1}(b).
First, a small upturn appears just above the superconducting transition at pressures up to 9.7 kbar, which seemingly evolves from the relatively flat resistance at 0 kbar. 
A similar upturn was observed in prior electrical transport measurements with current applied in the (011) plane, found to be accompanied by thermal hysteresis (not observed in this study) \cite{Ran2020}.
Second, the superconducting transition narrows and becomes sharpest at $p=9.7$ kbar, before broadening at higher pressures with a long tail just before the first-order transition to a magnetic phase occurs near $p=14.2$ kbar.
This feature was also observed previously  \cite{Ran2020}, and was shown to sharpen upon application of magnetic field. At higher pressures, the peak in $R(T)$ is diminished and a considerable increase in resistance occurs on cooling before an abrupt drop to finite resistance at the lowest measured temperatures. The features found above 15 kbar have been previously associated with magnetic ordering \cite{Li2021, Thomas2020}.

\begin{figure*}
\includegraphics[width=1\linewidth]{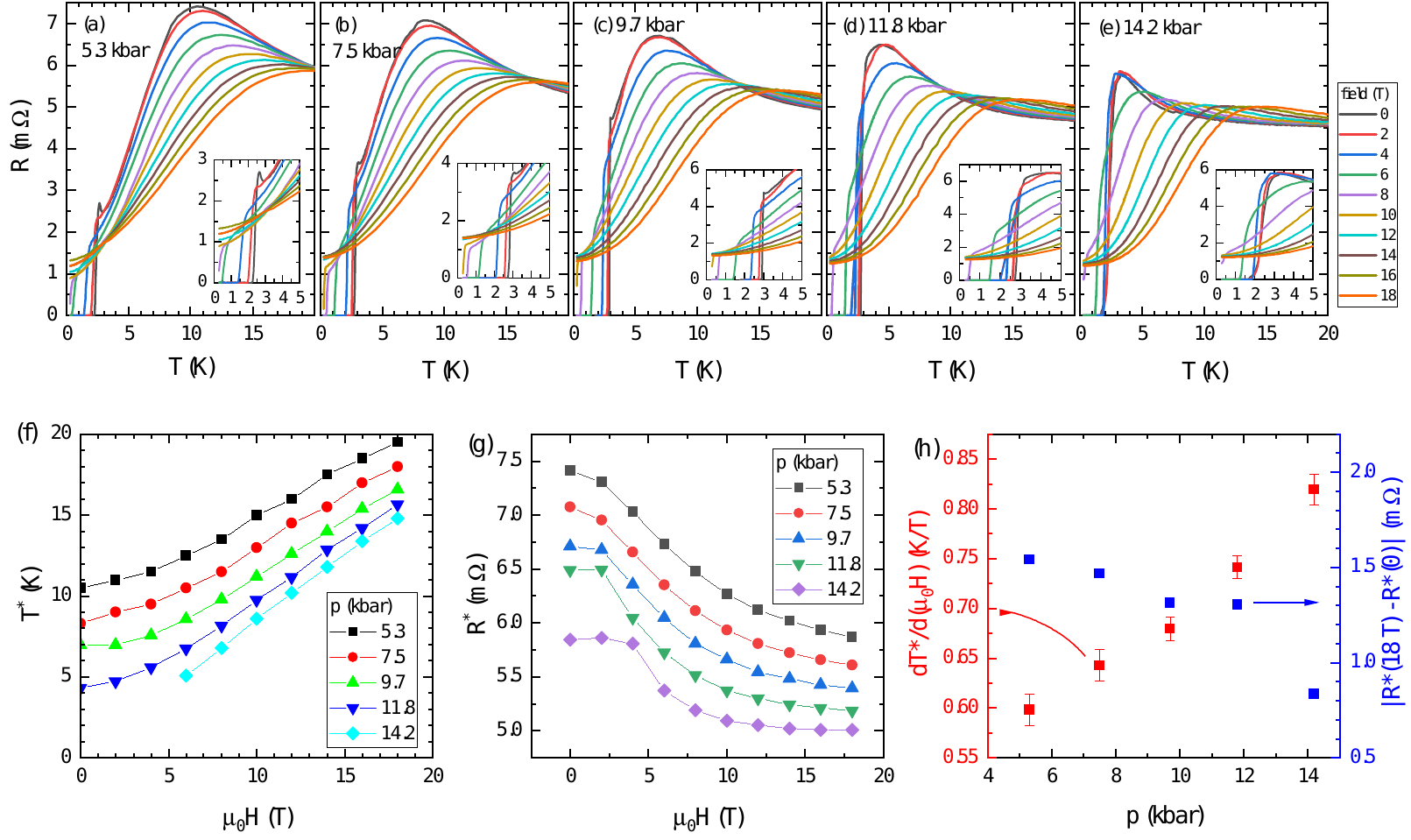}%
\caption{\label{fig2} 
{\bf Magnetic field evolution of $c$-axis resistivity of UT$_2$ under applied pressure.} 
Panels (a-e) show the field-evolution of $R(T)$ with applied pressure and fields applied along the $a$-axis, in the temperature range where the data exhibit a peak that evolves very sensitively with both pressure and magnetic field.  
Defining $T^*$ and $R^*$ as, respectively, the temperature and resistance at the maximum in $R(T)$, panels (f-g) show the field-dependence of these characteristic values to exhibit common features under all applied pressures. 
The pressure evolution of the rate of increase of $T^*$ with field, $dT^*/dH$, is plotted in panel (h) (left vertical axis), together with the total field variation of $R^*$, $|R^*(18~\textmd{T})-R^* (0)|$ (right vertical axis).}
\end{figure*}

The pressure-temperature phase diagram extracted from our $c$-axis resistivity measurements is presented in Figure~\ref{fig1}(c) as a contour plot, comparing the evolution of the resistivity magnitude with that of other measured quantities.
The precise resistivity measurements tracking the properties of the peak offer a clear picture, particularly near the critical pressure.
The zero-pressure $c$-axis peak at 13.8 K decreases in temperature with increasing pressure at a rate of $-$0.6 K/kbar, and the peak becomes narrower with pressure. 
Interestingly, the observed pressure suppression rate of the peak is in excellent agreement with that observed for the $a$-axis magnetic susceptibility  $\chi_a$, which is $-$0.58 K/kbar \cite{Li2021}.
Furthermore, Willa {\it et al.} \cite{Willa2021} estimated the pressure-suppression rate of the minimum thermal expansion coefficient along the $c$-axis from the thermodynamic Gr\"uneisen parameter to be $-$0.4 K/kbar, 
which also tracks the resistivity features as shown in Figure~\ref{fig1}(c).
Evidently, the pressure evolution of the $c$-axis peak closely tracks both the   $\chi_a$ feature as well as the Gr\"uneisen parameter, strongly suggesting all features have a common magnetic origin.

Applying magnetic field at each measured pressure reveals the field-evolution of $R(T)$ from 5.3 kbar to 14.2 kbar, where the $c$-axis peak remains as a pronounced local maximum but is strongly tuned by magnetic field. As shown in Figs.~\ref{fig2}(a-e), increasing magnetic field broadens the $c$-axis peak and increases its temperature position, while also invoking a shallower temperature dependence of the resistance with increased curvature. The broadening of the peak with field is similar to what was observed previously at ambient pressures \cite{Thebault2022}, but is contrary to the opposite trend observed with field applied along the magnetic hard axis ($b$-axis) \cite{Thebault2022,Knebel2023}.
To characterize this trend, we define $T^*$ and $R^*$ as, respectively, the temperature and resistance values at the $c$-axis peak for each pressure and field value, with the latter representing the field-evolution of the absolute low-temperature scattering rate at each pressure. 
The field-dependent $T^*$ and $R^*$ values show common features under all applied pressures with $H \parallel a$, as shown in Figs.~\ref{fig2}(f-g).
While $T^*$ increases with increasing field and approaches a linear trend, $R^*$ generically decreases with increasing field, except for a saturated evolution at low fields in the vicinity of the magnetic order transition.
The trends are characterized by plotting the rate $dT^*/d(\mu_0 H)$ (determined between 6 T and 18 T) and $|R^*(18~\textmd{T})-R^* (0)|$ in Fig. \ref{fig2}(h), which show nearly linear increase and decrease with pressure, respectively.

\begin{figure*}
\includegraphics[width=1\linewidth]{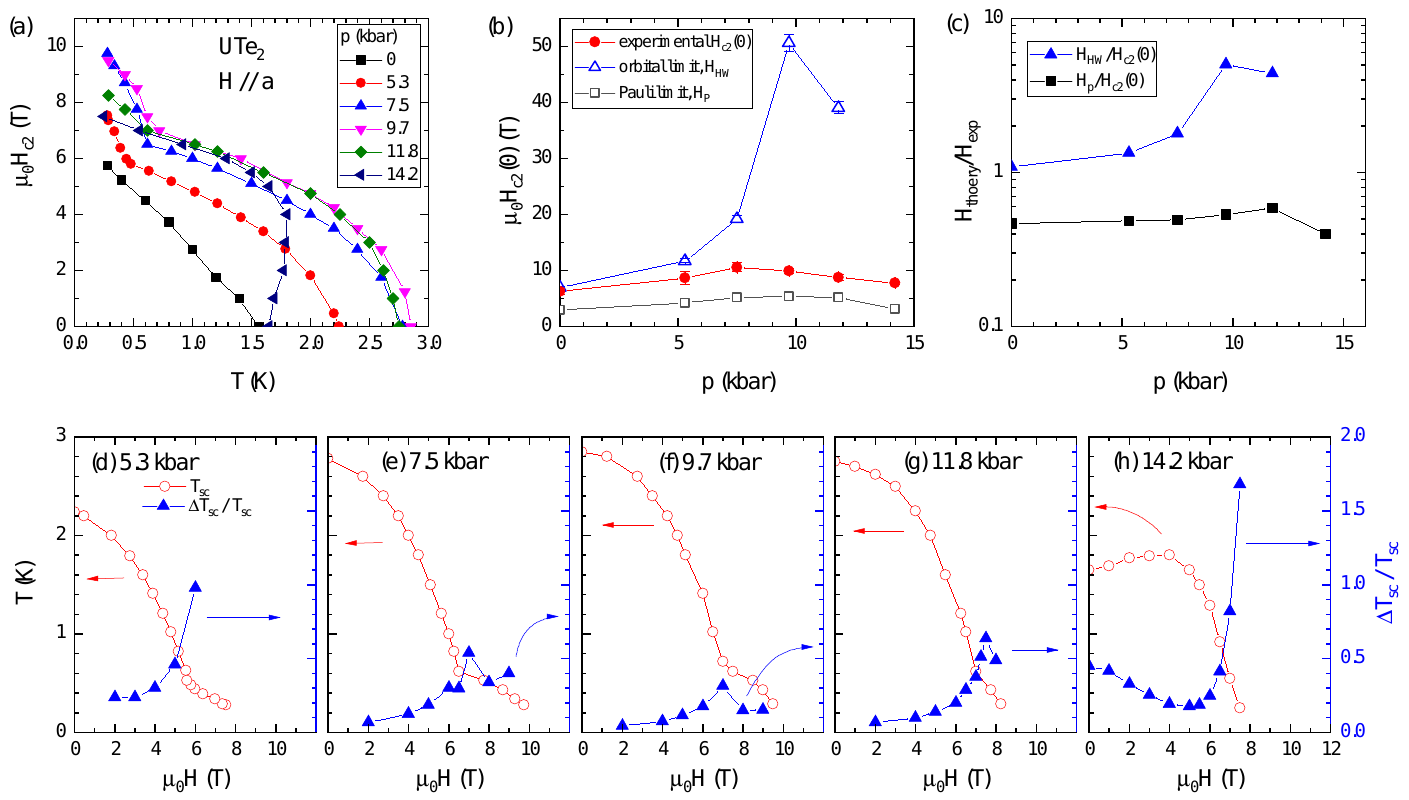}%
\caption{\label{fig3} {\bf Superconducting upper critical fields of UTe$_2$ as a function of applied pressure.} 
Panel (a) shows the temperature-dependent upper critical field $H_{c2}(T)$ under various applied pressures with field $H$ applied along the $a$-axis. Values are obtained using the zero resistance criteria for the superconducting transition temperature $T_{sc}$ in a magnetic field.
Panel (b) compares the extracted zero-temperature experimental $H_{c2}(0)$ values (red circles) to the calculated orbital limiting field, $H_{HW}$ (blue triangles), and the paramagnetic limiting field, $H_P$ (black squares). 
See text for definitions of $H_{HW}$ and $H_{P}$.
The experimental $H_{c2}(0)$ values are determined by extrapolating the $H_{c2}(T)$ curves to zero temperature.
Panel (c) shows the pressure evolution of $H_{HW}/H_{c2}(0)$ and $H_P/H_{c2}(0)$.
Panels (d-h) present the relation between the anomalous behavior $H_{c2}(T)$ (red circles) and the width of the superconducting phase transition $\Delta T_{sc}/T_{sc}$ (blue triangles). Under all measured pressures, the width exhibits strong enhancement in the field range where $H_{c2}(T)$ exhibits a sudden slope change, as discussed in the text.}
\end{figure*}

The effect of magnetic field on the superconducting transition also reveals interesting pressure evolution of the upper critical field $H_{c2}(T)$ as shown in Fig. \ref{fig3}.
The $H_{c2}(T)$ curves were determined from $R(T)$ measurements with the electrical current along the $c$-axis and the magnetic field applied parallel to the $a$-axis under applied pressure up to $p=14.2$ kbar.
We used the zero resistance criteria for the superconducting transition temperature $T_{sc}$ in field.
While the $H_{c2}(T)$ curve without the applied pressure exhibits a smooth variation, the application of pressure drastically changes the shape of the superconducting $H$-$T$ phase lines.
Near $T_c$, the slope of $H_{c2}(T)$ increases by almost five-fold under $p=9.7$ kbar, and it slightly decreases at 11.8 kbar, 
consistent with the previous results \cite{Aoki2020,Knebel2020}.
As was shown previously \cite{Ran2020}, the application of 14.2 kbar induces reentrant behavior of superconductivity.
The large slope change of $H_{c2}(T)$ at $T_c$  with pressure indicates the significant variation in the orbital limiting $H_{c2}(0)$ \cite{Helfand1966}. However, the overall observed $\mu_0 H_{c2}(T)$ at the lowest temperature remains between 6 and 10 T as shown in panel (a). 
When the field-driven superconducting to normal state transition occurs due to the orbital limiting effect, $H_{c2}(0)$ can be estimated from the slope of $H_{c2}(T)$ at $T_c$ with a relation, $H_{HW}=-\lambda T_c H'_{c2}(T_c)$, proposed by Helfand and Werthamer (HW) \cite{Helfand1966}. Here $\lambda\approx 0.73$ and 0.69, which correspond to the clean and dirty limits, respectively \cite{Helfand1966,Kogan2012}.
Alternatively, the spin-singlet superconductivity can be suppressed due to the Zeeman energy contribution of Pauli paramagnetism, and the limiting value $H_P$ can be estimated by a relation, 
$H_P=\Delta_0/\sqrt{2}\mu_B$. Here $\Delta_0$ and $\mu_B$ are the magnitudes of the superconducting energy gap at zero temperature and the Bohr magneton, respectively. For a weak-coupling BCS superconductor, $\mu_0 H_P=\alpha T_c$ where $\alpha\approx 1.87$ T/K.

\begin{figure*}
\includegraphics[width=1\linewidth]{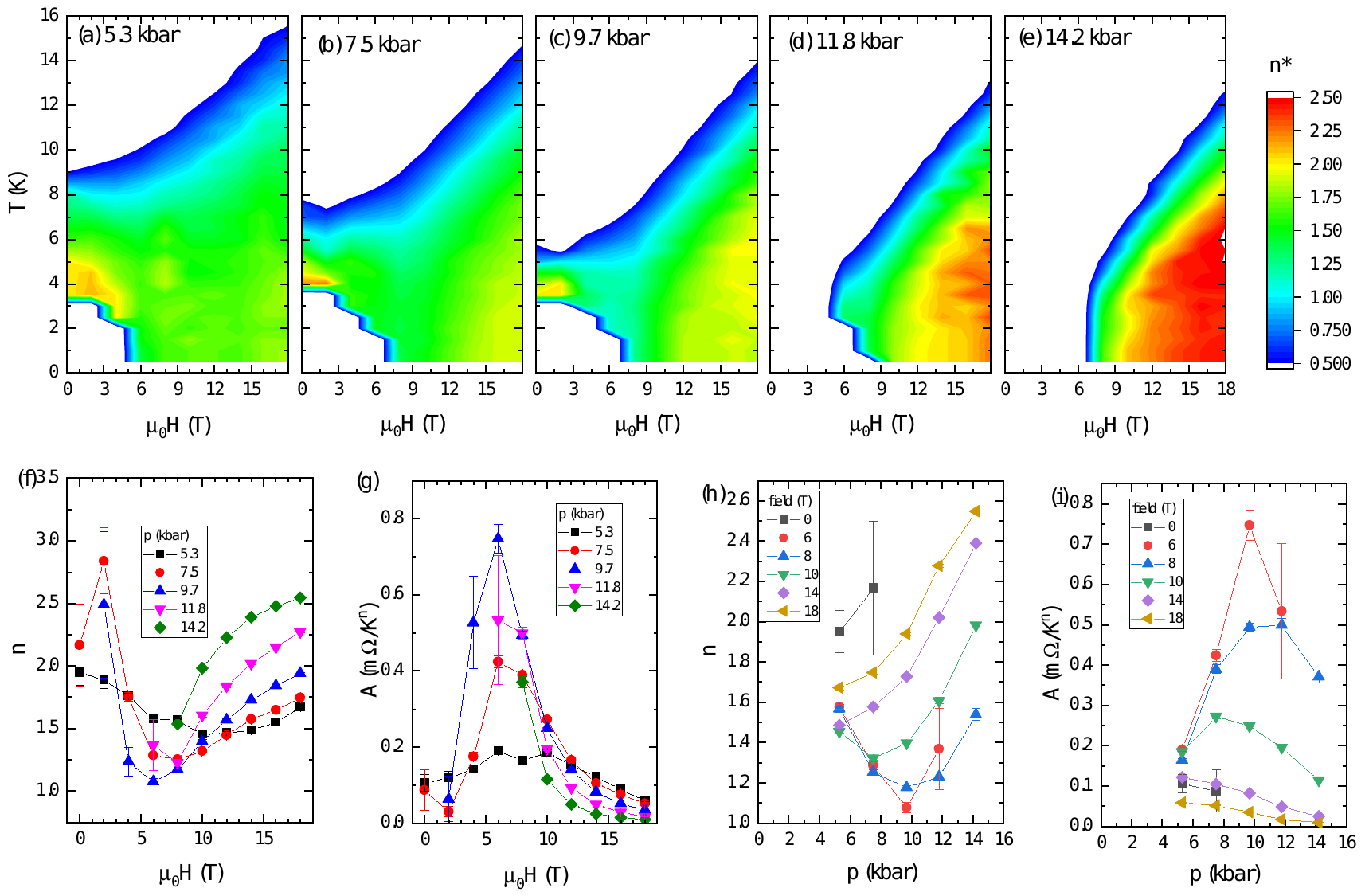}%
\caption{\label{fig4} 
{\bf Evolution of non-Fermi liquid behavior with field and pressure in UTe$_2$.}
Panels (a-e) show the field-dependent exponent $n^*$, representative of the power law exponent of the temperature dependence of $c$-axis resistance $R(T)$ for fields applied along the $a$-axis, determined using the relation $n^*=d[\log{(\rho(T)-\rho_0})]/d[\log{T}]$. 
At 5.3 kbar, $n^*$ exhibits Fermi Liquid behavior (i.e., $n^*=2$, shown as yellow coloring) just above $T_c$ near zero field, but decreases toward $n^*=1.5$ (light green) with increasing fields and decreasing temperatures. To quantify the trends, least-squares fitting of selected $R(T)$ using the relation $R(T) = R(0)+A T^n$ to the experimental data with $T \le T^*/2$ yield values for the extracted power law exponent $n$ and corresponding temperature coefficient $A$, summarized as a function of the field in panels (f,g) and pressure in panels (h,i).
}
\end{figure*}

We compare the experimental $H_{c2}(0)$ to both limiting fields, $H_{HW}$ and $H_P$, in Fig. \ref{fig3}(b). 
We note that $H_{HW}$ is ill-defined under $p=$ 14.2 kbar because of the reversed sign of $H'_{c2}(T_c)$, i.e., reentrant superconductivity.
Figure 3(c) shows the pressure evolution of $H_{HW}/H_{c2}(0)$ and $H_P/H_{c2}(0)$. 
While $H_P$ remains less than $H_{c2}(0)$, indicating non-singlet pairing, $H_{HW}$ exhibits a substantial variation.
The large $H_{HW}$ prediction is generally evidence for the heavy-fermion normal state \cite{Stewart1984}.
The pressure-evolution of $H_{HW}$, which exhibits a significant enhancement around 10 kbar, indicates increasing effective mass with pressure.
However, the orbital limiting effect is interrupted, and the largest discrepancy between $H_{c2}(0)$ and $H_{HW}$ is observed at 9.7 kbar where the highest $T_c$ is observed.
A similar effect was observed in other heavy fermion superconductors near quantum criticality \cite{Stewart1984}, suggesting the existence of a QCP near 10 kbar.
At low temperatures, a drastic slope change appears under pressure between 5.3 and 11.8 kbar.
The slope change in UTe$_2$ was previously reported by Aoki {\it et al.}, which was attributed to the existence of other superconducting phases \cite{Aoki2020}.
Similar $H_{c2}(T)$ behavior was reported by Kasahara {\it et al.} in FeSe \cite{Kasahara2020}, which was attributed to the Fulde--Ferrell--Larkin--Ovchinnikov (FFLO) state \cite{Fulde1964, Larkin1964, Matsuda2007}.

We found the width of the superconducting phase transition in resistivity is closely related to this anomalous behavior in $H_{c2}(T)$. To shed light on the origin of this feature, we determined the field-dependent transition width compared to the $T_{sc}$ that is determined at the zero resistivity, $\Delta T_{sc}/T_{sc}$. For all studied pressures, $\Delta T_{sc}/T_{sc}$ exhibits strong enhancement where the sudden slope change occurs as shown in panels (d-h). Defining $H^*$ as the field value where the slope of $H_{c2}(T)$ changes,
we observe that $\Delta T_{sc}/T_{sc}$ decreases above $H^*$ under $p=7.5$, 9.7, 11.8 kbar where the low-temperature data above $H^*$ are available.
A broad superconducting transition is usually associated with inhomogeneity \cite{Casalbuoni2004, Park2012, Thomas2021} or a filamentary superconducting state. However, the systematic field dependence rules out these simple scenarios, suggesting this is rather associated with the competing order parameters and quantum criticality leading anomalous transport properties.

Recently, the field evolution of the $c$-axis peak with $H \parallel a$ \cite{Thebault2022} and the pressure evolution of the $c$-axis transport with $H\parallel b$ \cite{Knebel2023} were reported. Here, we report the field and pressure evolution of the power-law temperature dependence of $\rho_c$ with field along the $a$-axis. Figs.~4(a-e) present the phase diagrams for each applied pressure determined by the field-dependent exponent $n^*$ of $R(T)$ determined using the relation $n^*=d[\log{(\rho(T)-\rho_0})]/d[\log{T}]$.
$R(0)$ is estimated by extrapolating the $R(T)$ tail assuming a power-law belavior of $R(T)$ in the low-temperature limit. Provided $R(0)$ is accurately determined, $n^*$ is equivalent to the exponent $n$, 
yielding a continuous approximate measure of the temperature power law exponent of $R(T)$. 
In previous reports, the $a$-axis resistivity of UTe$_2$ was shown to remain quadratic in temperature (i.e., $\Delta\rho_a \sim A T^n$, with $n$=2) for magnetic fields applied along both $a$- and $b$-axes up to 40~T, with the coefficient $A$ significantly enhanced near a 35 T $b$-axis field \cite{Knafo2019} but retaining Fermi liquid (FL) behavior.
Linear in temperature (i.e., $n=1$)  resistivity was reported by Thomas {\it et al.} \cite{Thomas2020} in the $a$-axis transport at low temperatures around 13 kbar.
For $c$-axis resistivity, Eo {\it et al.} reported quadratic FL behavior in the absence of both field and applied pressures \cite{Eo2022}. 
As shown in Fig.4, $R(T)$ exhibits FL behavior (yellow) just above $T_c$ at $p=5.3$ kbar in zero field, but the exponent $n^*$ decreases toward $n^*=1.5$ (light green) with increasing field near $H_{c2}(0)$.
Under 7.5 kbar and 9.7 kbar, while the $c$-axis transport exhibits non-FL behavior near $H_{c2}(0)$,  FL behavior (yellow) is recovered at high fields between 15 T and 18 T.
Under 11.8 kbar and 14.2 kbar, the exponent reaches $n^*=2.5$ (red) at high fields.

Whereas the FL behavior (i.e., $T^2$) is expected at low temperatures in a typical metal, a non-FL sub-quadratic exponent is a telltale signature of unconventional scattering that has been attributed to the presence of enhanced spin fluctuations near a magnetic quantum critical point \cite{Stewart1984, Mathur1998, Paglione2003}.
To study the quantitative trends, we performed least-squares fitting on selected $R(T)$ curves by fitting our data to the relation $R(T) = R(0)+AT^n$ with $T \le T^*/2$. 
The field evolution of $n$ and $A$ are summarized in panels (f, g) and pressure evolution in panels (h, i).
For $p=5.3$ kbar, $n=2$ in zero field but smoothly decreases with increasing field, showing a minimum value of $n=1.5$ near 10 T. It weakly increases at high fields while remaining sub-quadratic up to the highest fields measured.
For higher pressures between 7.5 and 11.8 kbar, $n$ exhibits a more drastic decrease with a minimum near 6-8 T where the $H_{c2}(T)$ changes the slope.
The smallest exponent $n\approx 1$ is observed near 6 T under 9.7 kbar. 
At higher fields, $n$ increases substantially to about 2.5 for 11 kbar and 14.2 kbar.
The extracted $A$-coefficient appears to correlate inversely with the trends in the power law exponent, with a dip in $n$ and a peak in $A$ at a field near the suppression of the superconducting state being typical for a system at or near a quantum critical point. 
In UTe$_2$, this signature in $c$-axis transport is a revealing indication of an incipient magnetic order that has a strong influence on the physical properties and possibly the superconductivity.

\section{Methods}

{\bf Sample preparation:} Single crystals of UTe$_2$ were synthesized by the chemical vapor transport method using iodine as the transport agent. 
Elements of U and Te with atomic ratio 2:3 were sealed in an evacuated quartz tube, together with 3 mg/cm$^3$ iodine. The ampoule was gradually heated up and held in the temperature gradient of 1060/1000 \degree C for 7 days, after which it was furnace cooled to room temperature.


{\bf Transport measurements under pressure:} A UTe$_2$ single-crystal sample with an onset $T_c\approx 1.78$ K was prepared for transport measurements by soldering electrical leads with gold wires.
The typical contact resistance is less than 1 $\Omega$.
The transport data were taken with a fixed current of 100 $\mu$A.
A nonmagnetic piston-cylinder pressure cell was used for measurements under pressure up to 17.4 kbar, with Daphne oil 7373 as the pressure medium. 
Transport measurements were performed in a commercial $^3$He cryostat system with a base temperature of 300 mK, which is equipped with a superconducting magnet. The current was applied along the crystallographic $c$-axis. 
The magnetic field up to 18 T was applied along the $a$-axis, perpendicular to the current. 
The pressure produced on the single-crystal sample at low temperatures was calibrated by measuring the superconducting transition temperature of lead placed in the cell. The known pressure dependencies of the superconducting transition temperature of Pb \cite{Smith1967, Ran2020} were used for this purpose.

\section{acknowledgments}
\begin{acknowledgments}
The authors are grateful for the useful discussions with Andriy Nevidomskyy.
Research at the University of Maryland was supported by the Department of Energy Award No. DE-SC-0019154 (transport experiments), the Gordon and Betty Moore Foundation’s EPiQS Initiative through Grant No. GBMF9071 (materials synthesis), NIST, and the Maryland Quantum Materials Center.

\end{acknowledgments}

\end{document}